\documentstyle[psfig]{aa}

\begin{document}

\title{Infrared imaging and spectroscopy of companion candidates near 
the young stars HD 199143 and HD 358623 in Capricornius \thanks{Based on 
observations obtained on La Silla,
Chile, in ESO programs 66.D-0135, 67.C-0209, 67.C-0213, 68.C-0009.} }

\author{R. Neuh\"auser\inst{1}, E. Guenther\inst{2}, M. Mugrauer\inst{1}, Th. Ott\inst{1},
A. Eckart\inst{3}}

\offprints{R. Neuh\"auser, rne@mpe.mpg.de }

\institute{MPI f\"ur extraterrestrische Physik, Giessenbachstra\ss e 1, D-85740 Garching, Germany
\and Th\"uringer Landessternwarte Tautenburg, Sternwarte 5, D-07778 Tautenburg, Germany
\and I. Physikalisches Institut, Universit\"at zu K\"oln, Germany}

\authorrunning{Neuh\"auser et al.}
\titlerunning{Companion candidates near HD 199143 and HD 358623}

\date{Received May 2002 / Accepted Sept 2002}

\abstract{We present JHK images of the young ($\sim 20$ Myrs) nearby ($\sim 48$ pc)
stars HD 199143 and HD 358623 (van den Ancker et al. 2000) 
with high sensitivity and high dynamic range 
in order to search for (sub-)stellar companions around them.
The images were obtained in JHK with the speckle camera SHARP-I 
in July 2001 and in H with the infrared imaging camera SofI 
in December 2000, July 2001, and December 2001, all at the ESO 3.5m NTT.
We present a companion candidate 
with a 2 arcsecond offset being
almost 2 mag fainter than HD 358623
with proper motion (over one year baseline) consistent with the 
known proper motion of the primary star HD 358623 A and $6 \sigma$ 
deviant from the assumption that it is a non-moving background object.
Then, we obtained a spectrum in the H-band (with SofI at the NTT) 
of this companion showing that it has spectral type 
M2 ($\pm 1$),
consistent with its JHK colors (for negligible extinction) and with
being a companion (i.e. at same age and distance) of HD 358623 A (K7-M0),
given the magnitude difference.
Also, a companion candidate 
with a 1 arcsecond offset being
2 mag fainter than HD 199143
is detected, but clearly resolved from the primary only with SHARP-I,
so that we have no proper motion information. Also, we could not obtain
a spectrum with SofI due to the high dynamic range.
The JHK colors of this candidates and the magnitude difference between 
primary (F8) and companion candidate are consistent with a spectral type M0-2.
This companion candidate was predicted by van den Ancker et al. (2000) 
to explain some unusual properties of the primary star HD 199143 A.
We determine the ages and masses of all four objects from theoretical
tracks and isochrones, all four stars appear to be co-eval with $\sim 20$ Myrs.
Both these companions were presented previously by Jayawardhana \& Brandeker (2001),
but only single-epoch images in J and K obtained in May 2001.
Our limits for additional detectable but undetected companions are such that we
would have detected all stellar companions with 
separations $\ge 0.5^{\prime \prime}$ (24 AU at 48 pc). 
}

\maketitle

\section{Introduction: HD 199143 and HD 358623}

HD 199143 and HD 358623 were presented as young, nearby, co-moving stars
by van den Ancker et al. (2000; henceforth vdA00). HD 199143 is at a distance
of $47.7 \pm 2.4$~pc as measured by Hipparcos.
Because HD 199143 and HD 358623 are only five arc min apart from each other 
and both show activity (a youth indicator) and the same proper motion, vdA00 suggested 
that they form a (small, but possibly larger) young nearby moving group (Capricornius),
similar to the HorA, Tuc, TW Hya, and $\beta$ Pic moving groups.
HD 199143 has similar $UVW$ space motion as the Tuc and TW Hya stars
(vdA00; Zuckerman \& Webb 2000). Four more member candidates were presented 
by van den Ancker et al. (2001; henceforth vdA01) selected within $5^{\circ}$ 
around HD 199143 by strong ROSAT X-ray emission and, partly, by proper motion.
It would be very important to find more members to this new association,
not only to study the formation and mass function of these new young
nearby groups, but also because such young nearby stars are very well
suited for direct imaging searches of sub-stellar companions.

HD 199143 (also called BD$-17^{\circ}6127$ and SAO 163989)
has spectral type F8 (Houk \& Smith-Moore 1988; vdA00)
and displays anomalous ultra-violett emission, Ca H \& K emission, and fast rotation (vdA00).
Also, HD 199143 shows N- and Q-band excess (vdA01) as well as 
IRAS $12 \mu m$ excess emission, but very low upper limits in the other IRAS bands 
(vdA00), so that this star shows moderate, but not strong infrared (IR) excess 
emission. However, vdA00 argue that all its features could be explained by a faint
T-Tauri-like companion, which produces itself the UV and IR excess emission
and whose circumstellar material accretes onto HD 199143 and thereby spins up the primary.
Mora et al. (2001) measured a rotational velocity of $v \cdot \sin i = 155 \pm 8$ km/s.

HD 358623 (also called BD$-17^{\circ}6128$ and AZ Cap) has spectral type K7-M0
with H$\alpha$ emission, strong flaring, and strong Li 6708\AA~absorption
(Mathioudakis et al. 1995; vdA00), as well as N- and Q-band excess 
(vdA01), all typical for a T~Tauri star. The Tycho proper motion of
HD 358623 ($\mu _{\alpha} = 59 \pm 3$ mas/yr and $\mu _{\delta} = -63 \pm 3$ mas/yr)
is identical to that of
HD 199143 ($\mu _{\alpha} = 59.2 \pm 1.1$ mas/yr and $\mu _{\delta} = -61.55 \pm 0.85$ mas/yr).
Recently, Zuckerman et al. (2001) argued that both HD 199143 and HD 358623
are actually part of the $\beta$ Pic moving group.

In May 2001, Jayawardhana \& Brandeker (2001; henceforth JB01) observed the two stars  
in J \& K with the AO system ADONIS at the ESO 3.6m telescope on La Silla.
Near each of the two stars, they detected one companion candidate. The close and
faint object near HD 199143 is red ($J-K=1.4$ mag) and, hence, either sub-stellar
or the (reddened) companion expected by vdA00, whose circumstellar material could 
explain the anomalous features of the primary HD 199143.
The companion candidate near HD 358623 is less than 2 mag fainter than the primary
in J \& K (JB01) and could be an M-type stellar companion.

We observed the two stars with high sensitivity and high dynamic range 
in order to detect faint companions, namely in the H-band 
in Dec. 2000, July 2001, and Dec. 2001 and in the J- and K-bands in July 2001
(speckle and normal IR imaging).
Young nearby stars like HD 199143 and HD 358623 are well-suited for direct imaging
of sub-stellar companions, both brown dwarfs and giant planets, because young
sub-stellar objects are still relatively bright (e.g. Burrows et al. 1997),
so that they are less difficult to detect in the PSF wing of a much brighter star.
See, e.g., Lowrance et al. (1999, 2000), Neuh\"auser et al. (2000b), and
Guenther et al. (2001) for imaging and spectroscopy of brown dwarf companions
of the young stars TWA-5 and HR 7329 in the TW Hya and Tuc associations.
The brown dwarf near TWA-5 was the first sub-stellar companion around a
pre-main sequence star confirmed by both proper motion and spectroscopy,
and also the first around a spectroscopic binary (Torres et al. 2001),
and the first around a star with evidence for a disk (Jayawardhana et al. 1999).
While the first four brown dwarfs confirmed as companions all orbit M-type stars, 
an A-type star also can have a brown dwarf companion (e.g. HR 7329 with spectral type A0). 
Therefore, both HD 199143 (F8) and HD 358623 (K7-M0) are promising 
targets for the direct imaging search for sub-stellar companions.

The probability for the two companion candidates detected by JB01 to be unrelated
background objects happen to lie in the line-of-sight next to the primary stars
is very small (JB01). However, one should not rely on such probabilities,
even when observing only a small sample. Some previous very faint sub-stellar 
companion candidates (e.g. Terebey et al. 1998, Neuh\"auser et al. 2000a)
with very low background probability were found to be background stars
by follow-up spectroscopy (Terebey et al. 2000, Neuh\"auser et al. 2001).
This shows how important it is to take multi-epoch images and spectra.

We present our imaging observations in Sect. 2 and the resulting
photometry for the two stars and their companion candidates in Sect. 3.
Astrometry is presented in Sect. 4 to check whether the companion candidates
are co-moving with their putative primary stars. Then, in Sect. 5, 
we present an H-band spectrum of one of the two companion candidates. 
We conclude in Sect. 6.

\section{Imaging observations}

We observed HD 199143 and HD 358623 
several times
with the Son of Isaac
(SofI\footnote{see www.ls.eso.org/lasilla/Telescopes/NEWNTT/}) at the 3.5m New 
Technology Telescope (NTT) of the European Southern Observatory (ESO) on La Silla, Chile.
%
%
The SofI detector is an Hawaii HgCdTe $1024 \times 1024$ array with $18.5 \mu$m pixel sizes.
We used the small SofI field with its best pixel scale for better angular resolution 
and determined the pixel scale by comparing the separations between several stars on other
images taken in the same night with 2MASS images of the same fields to be 
$0.150 \pm 0.002^{\prime \prime}$ per pixel.
Darks, flats, and standards were observed in the same nights with the
same set-up and data reduction was done with
{\em eclipse}\footnote{see www.eso.org/projects/aot/eclipse/}
version 3.8, a C-based software library. While {\em eclipse} is made for VLT data 
reduction, like e.g. the Infrared Imaging And Array Camera (ISAAC), and not 
guaranteed to work for SofI data, it also does work for SofI imaging data 
reduction (dark, flat, shift+add); after all, SofI is the Son of Isaac.
See Table 1 for the observations log.

\begin{table}
\begin{tabular}{lcccl} 
\multicolumn{5}{c}{\bf Table 1. NTT observations log} \\ \hline
Instr. & obs. date & exp. [s] & band & FWHM \\ \hline
\multicolumn{5}{c}{HD 199143 at $\alpha = 20:55:47.67$ and $\delta = -17:06:51.0$} \\ \hline
SofI    & 7 Dec 2000 & $560 \times 1.5$ & H & 0.78$^{\prime \prime}$ \\
Sharp-I & 1 Jul 2001 & $400 \times 0.5$ & J & 0.56$^{\prime \prime}$ \\
Sharp-I & 1 Jul 2001 & $400 \times 0.5$ & H & 0.49$^{\prime \prime}$ \\
Sharp-I & 1 Jul 2001 & $1200 \times 0.5$ & K & 0.40$^{\prime \prime}$ \\
SofI    & 8 Jul 2001 & $400 \times 1.3$ & H & 1.03$^{\prime \prime}$ (*)\\
SofI    & 7 Dec 2001 & $400 \times 1.5$ & H & 1.02$^{\prime \prime}$ \\ \hline \hline
\multicolumn{5}{c}{HD 358623 at $\alpha = 20:56:02.77$ and $\delta = -17:10:54.1$} \\ \hline
SofI    & 7 Dec 2000 & $200 \times 1.2$ & H & 0.88$^{\prime \prime}$ \\
Sharp-I & 2 Jul 2001 & $400 \times 0.5$ & J & 0.54$^{\prime \prime}$ \\
Sharp-I & 2 Jul 2001 & $400 \times 0.5$ & H & 0.41$^{\prime \prime}$ \\
Sharp-I & 2 Jul 2001 & $1500 \times 0.5$ & K & 0.36$^{\prime \prime}$ \\
SofI    & 6 Dec 2001 & $500 \times 1.2$ & H & 0.79$^{\prime \prime}$ \\ \hline
\end{tabular}
Remarks: Positions given for J2000.0. (*) Not photometric.
\end{table}

\begin{figure}
\vbox{\psfig{figure=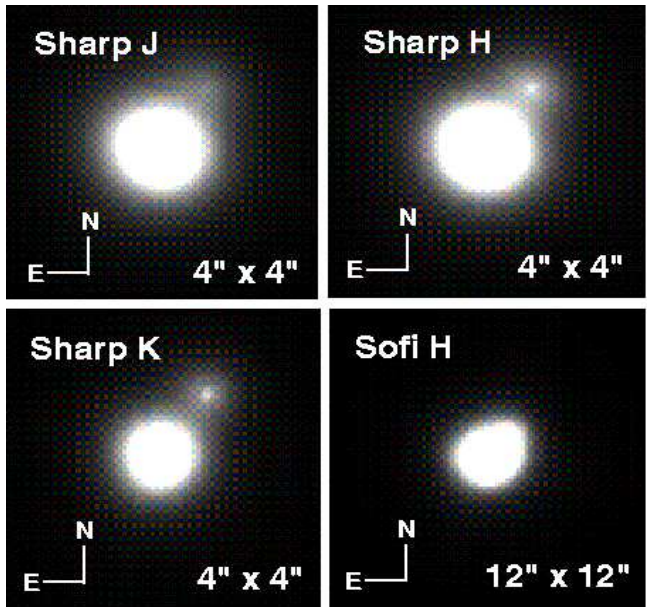,width=8.5cm,height=8cm}}
\caption{HD 199143 and its faint close companion candidate in our
SHARP-I images in JHK and our SofI H-band image.
The lower right panel (SofI) shows a larger field, where no
additional companion candidates are detected.
}
\end{figure}

\begin{figure}
\vbox{\psfig{figure=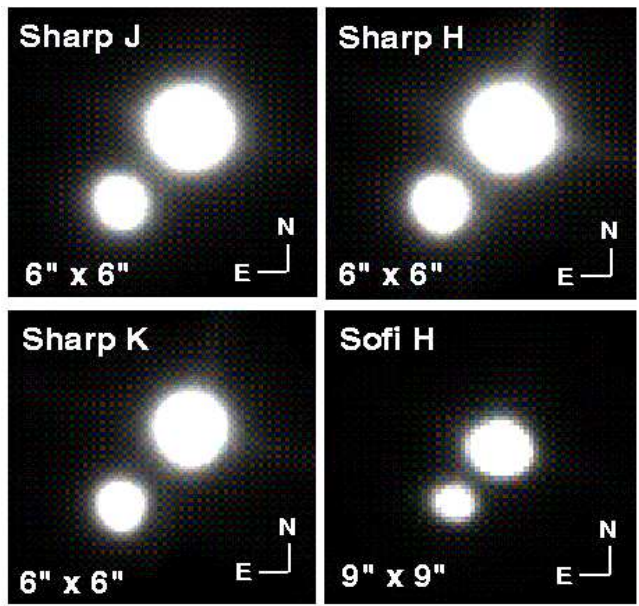,width=8.5cm,height=8cm}}
\caption{HD 358623 and its two companion candidates in our
SHARP-I images in JHK and our SofI H-band image.
The lower right panel (SofI) shows a larger field, where no
additional companion candidates are detected.
}
\end{figure}

Then, we observed HD 199143 and HD 358623 at the end of the nights of
1 \& 2 July 2001, respectively, using SHARP-I (System for High Angular
Resolution Pictures, Hofmann et al. 1992) at the NTT.
%
%
The targets were placed onto the two lower, i.e. western SHARP-I quadrants,
because they have better pixel and flat field characteristics.
The data were corrected for bad pixels followed by a sky image subtraction and
the application of a flat-field. For each band we then co-added the $256 \times 256$
pixel frames using the brightest pixel as shift-and-add reference (Christou 1991).
Exposure times and FWHM in the final co-added images (using the SHARP 
pixel scale of $0.0491^{\prime \prime}$, see also Neuh\"auser et al. 2000a) 
are given in Table 1.

%
%
%
%

\section{Companion candidates: Photometry}

The companion candidates detected in J \& K by JB01 in their May 2001 ADONIS images
are also detected in our H-band SofI images in Dec. 2000 and Dec. 2001 and
as well in JHK in our July 2001 SHARP-I images.
We obtained the JHK magnitudes of the stars and companion candidates from the
SHARP-I images (because the objects are best separated on those image, see Fig. 1 \& 2)
by normal aperture photometry with an aperture size just small enough to exclude 
the companion, compared to the 
flux standard HR 8278 (van der Bliek et al. 1996)
observed in JHK immediately after the respective science target. 
Then, the magnitudes for the companion candidates are obtained 
from the background-subtracted flux ratio with their respective primary.
The resulting photometric data are listed in Table 2.

\subsection{HD 199143}

We display the SHARP-I JHK images of HD 199143 and its close ($1^{\prime \prime}$)
companion candidate in Fig. 1 together with the Dec. 2000 SofI image,
where the companion candidate (called HD 199143 B in JB01)
was marginally detected.

\begin{table}
\begin{tabular}{lccccc}
\multicolumn{6}{c}{\bf Table 2. Photometry} \\ \hline
Object      & J    & H    & K    & A$_{\rm V}$ & SpecType \\ \hline
HD 199143 A & 6.2  & 6.0  & 6.0  & $\le 0.1$ & F8 (1,2) \\
HD 199143 B & 9.2  & 8.4  & 8.2  & $\le 0.1$ & M0-2 (3) \\ \hline
HD 358623 A & 8.0  & 7.3  & 7.2  & $\le 0.1$ & K7-M0 (1,4) \\
HD 358623 B & 9.7  & 9.0  & 8.8  & $\le 0.1$ & M$2 \pm 1$ (5) \\ \hline
\end{tabular}
Magnitudes and absorptions from our data ($\pm 0.1$ mag).
References for spectral types: (1) Simbad, (2) vdA00, 
(3) this work, from JHK photometry (Sect. 3.2),
(4) Mathioudakis et al. (1995), 
(5) this work, from an H-band spectrum (Sect. 5).
\end{table}

Our data on HD 199143 A agree well with those in JB01, within our 
more conservative precision of $\pm 0.1$ mag (as we observed less standards),
while HD 199143 B is 0.3 mag brighter in our data than in JB01,
possibly due to problems in JB01 with the subtraction of the bright background 
due to the PSF of HD 199143 A (R. Jayawardhana, priv. comm.).

The IR colors $J-H=0.2 \pm 0.1$ and $H-K=0.0 \pm 0.1$ mag (Table 2) for HD 199143 A 
are consistent with its spectral type F8 and its optical colors, for negligible 
foreground extinction. 
The companion candidate to HD 199143 is redder 
than the primary,
either because it is less massive and/or more reddened, namely by either 
circumstellar material and/or other foreground extinction. 
In case of negligible absorption, as towards the primary, the JHK colors
of the companion candidate would be consistent within the errors with a mid-K
to mid-M type dwarf. If the companion candidate is at the same distance as 
the primary and also on the zero-age main-sequence, than its absolute 
JHK magnitudes are best consistent with an early M-type dwarf. 
Hence, our spectral type classification is M0-2.

\subsection{HD 358623}

HD 358623 and its close ($2^{\prime \prime}$) companion candidate 
(called HD 358623 B in JB01) are shown in Fig. 2.
The primary star and the companion candidate are also well separated and detected in our 
SofI images, so that we can obtain photometric and astrometric data from both 
the SofI and the SHARP-I images.
In case of the SofI images, we used the faint HST standard stars S-361-D
and S-754-C (Persson et al. 1998), and the data agree well with our 
SHARP-I H-band magnitudes.
Our data on HD 358623 A and B also agree well with those in JB01.
Hence, we have no indication for variability.

The IR colors $J-H=0.7 \pm 0.1$ and $H-K=0.1 \pm 0.1$ mag for HD 358623 A are 
consistent with its spectral type K7-M0, for negligible extinction.
There is no reddening observed in the near IR in these two stars, consistent with 
the spectral energy distribution shown in vdA01, where some
moderate IR excess is seen only in the thermal IR.

The IR colors of the companion candidate HD 358623 B are very similar as for the primary
HD 358623, but also consistent with any spectral type between mid-K and mid-M.
Furthermore, the companion candidate is only slightly fainter (less than 2 mag)
than the primary, so that it could be a slightly redder stellar secondary with 
slightly less mass and slightly later spectral type 
(early to mid-M).
A spectral type M0-3 is best consistent with its absolute JHK-band magnitudes,
when assuming that is has the same distance as HD 199143 A and that
it is also located on the zero-age main-sequence.
This classification will be refined by an H-band spectrum (Sect. 5).

Whether the two companion candidates are truely bound companions or (possibly reddened) 
background stars can be found out best by spectroscopy and proper motion.

\section{Astrometry: Common proper motion pairs~?}

%
%

Because we have observed both HD 199143 and HD 358623 several times (Dec. 2000,
July 2001, Dec. 2001), we can check, whether the companion candidates share
the proper motion with their respective primaries.
In the case of HD 358623, we have detected the companion candidate
at all three epochs, while in the case of HD 199143, the companion candidate
is only marginally resolved in the two SofI images. We could compare our
separations between HD 199143 A and B with those given in JB01, but that does not
yield any significant conclusion, because the JB01 observations were obtained
only one month before our SHARP-I images.
%
%

\begin{table*}
\begin{tabular}{lcccccc}
\multicolumn{7}{c}{{\bf Table 3. Separations of the companion candidates} ($\pm 1 \sigma$ errors)} \\ \hline
Pair & $\Delta \alpha [^{\prime \prime}]$ & $\Delta \delta [^{\prime \prime}]$ & sep. $[^{\prime \prime}]$ & PA [$^{\circ}$] & epoch & Ref. \\ \hline
HD 199143 A \& B & $0.619 \pm 0.016$ & $0.884 \pm 0.015$ & $1.079 \pm 0.009$ & $325.0 \pm 0.7$ & 2001.4 & JB01 \\
HD 199143 A \& B & $0.595 \pm 0.024$ & $0.832 \pm 0.019$ & $1.023 \pm 0.031$ & $324.43 \pm 1.60$ & 2001.5 & this work \\ \hline
HD 358623 A \& B & $1.433 \pm 0.002$ & $1.661 \pm 0.008$ & $2.194 \pm 0.011$ & $139.21 \pm 0.68$ & 2000.9 & this work \\
HD 358623 A \& B & $1.434 \pm 0.019$ & $1.673 \pm 0.018$ & $2.203 \pm 0.007$ & $139.4 \pm 0.5$ & 2001.4 & JB01 \\
HD 358623 A \& B & $1.426 \pm 0.013$ & $1.640 \pm 0.015$ & $2.178 \pm 0.007$ & $138.99 \pm 1.02$ & 2001.5 & this work \\
HD 358623 A \& B & $1.433 \pm 0.011$ & $1.725 \pm 0.010$ & $2.243 \pm 0.021$ & $140.28 \pm 0.88$ & 2001.9 & this work \\ \hline
\end{tabular}
\end{table*}

\begin{figure}
\vspace{-1cm}
\vbox{\psfig{figure=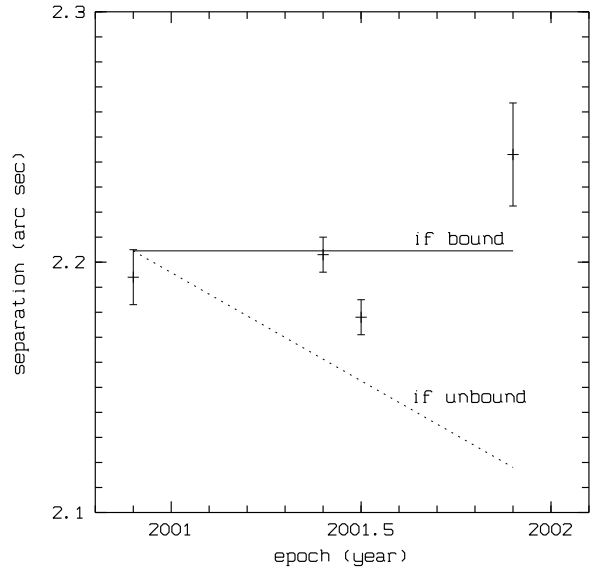,width=13cm,height=9cm,angle=270}}
\caption{Measured separation (in arc sec, 
with $1 \sigma$ error bars)
versus observing date (in years) between HD 358623 A and B with
data points from Table 3. 
The solid line show the median separation.
If the visual pair really is a bound pair, then the separation should not change
(except for orbital motion, which is negligible for our relatively
wide separation and small epoch difference). If the 
two stars were unrelated,
then the separation should change according to the
proper motions of the two stars, i.e. mainly
the known proper motion of the primary star,
assuming that the companion candidate is a 
non-moving background star; the changes in separation expected for that 
case are shown as dashed line. 
For the pair HD 358623 A \& B, we have significant evidence for 
common proper motion:
The companion is located SE of the star and the star
is known to move to the SE, so that the observed separations
should have decreased, if the pair would have been unbound;
the last data point is $6 \sigma$ deviant from the background hypothesis,
but less than $2 \sigma$ deviant from the companion hypothesis.
}
\end{figure}

The separations are measured as differences between the photo-centers of the
objects determined by a Gaussian centering
with {\em MIDAS}~\footnote{see www.eso.org/projects/esomidas/},
with the pixel scales mentioned above.
The north-south alignment of SHARP-I during our observing nights 
was determined to be precise within less than half a degree using images
of the galactic center and known bound binaries taken in the same nights.
The separations in $\alpha$ and $\delta$ as well as the position angles PA
given in Table 3 include an additional error of $\pm 0.5^{\circ}$ 
due to an uncertainty in the north-south alignment of the detectors,
while the errors in total separation are independant of that uncertainty.

First, we have measured 18 other comparison stars in the HD 358623
field, in order to determine the precison of our astrometry.
%
%
The separations between each pair of comparison stars did not change
within the 10 to 20 mas errors.
Hence, they are very likely non-moving background stars.
%
%
HD 358623 moves relative to the comparison stars to the SE. 
From the difference in the separations between 18 comparison stars
and HD 358623, we can measure its proper motion ($\sim 103$ mas/yr), 
which is consistent with its known Tycho proper motion ($\sim 86$ mas/yr)
within less than 20 mas/yr.
This shows how precise relative astrometry can be, even after only one
year of epoch difference, due to (i) the small SofI pixels
together with the small FWHM due to excellent seeing
and active optics (hence small error in Gaussian centering),
(ii) the relatively large SofI field-of-view even in the
so-called {\em small SofI field}
(hence many sufficiently bright comparison stars),
(iii) and the stability of the telescope and instrument optics
over the one year epoch difference.

%
%

If the close companion candidate would be an unrelated background object, it should also
move relative to HD 358623 as the other very widely separated background objects,
namely by almost $\sim 100$ mas/yr according to the PA.
The separations measured by us and by JB01 are given in Table 3.
One can clearly see that the companion candidate HD 358623 B did not
move relative HD 358623 A 
(see Fig. 3)
from one epoch to any of the three other epochs,
not within the measurement errors and certainly not with a motion as large
as $\sim 100$ mas/yr. In particular, the comparison between the Dec. 2000 SofI image
and the Dec. 2001 SofI image is very significant, because they have been taken
with the same instrument at the same airmass and also with exactly a full year 
epoch difference, so that there are no differential parallax nor refraction effects.
Because the companion candidate is located SE of the primary and because the
primary is known to move SE, the separation should have decreased by $\sim 100$ mas, 
if the pair would not be bound. This can be excluded with a significance of $6 \sigma$.
Hence, the companion candidate is co-moving with the star. Therefore, it is
very likely a truely bound companion of the young nearby star HD 358623.
To confirm this conclusion, we have taken a spectrum of HD 358623 B.

\section{Spectrocopy of HD 358623 A and B}

%
%
The spectral type of HD 358623 B was determined in Sect. 3.2 roughly to be
early- to mid-M from its JHK magnitudes.
To confirm and refine this classification,
we obtained an H-band spectrum on 8 Dec 2001 with SofI.
%
%
We took 20 spectra with 60 sec exposure each through a 
$1^{\prime \prime}$ slit with a red grism including both the H- and K-band 
(1.53 to $2.52~\mu$m) with a resolution of $R \simeq 1000$.
Data reduction was done in the normal way with IRAF:
Dark subtraction, normalization, flat fielding, sky subtraction, 
wavelength calibration, co-adding the spectra, then
correction for instrumental sensitivity and atmospheric response.
The spectra were not flux-calibrated.
The final spectra of both the primary HD 358623 A
and the secondary HD 358623 B are shown in Fig. 4.
Because we have comparison spectra of K- to M-type dwarfs 
available only for the H-band, we use and show only the H-band 
part of the SofI H+K-band spectrum.
The H-band spectrum of the comparison star GSC~7210~1352, a known M1-dwarf
in the TW Hya region, has been taken by us with ISAAC at the VLT
and was published before in Neuh\"auser et al. (2002).

\begin{figure}
\vbox{\psfig{figure=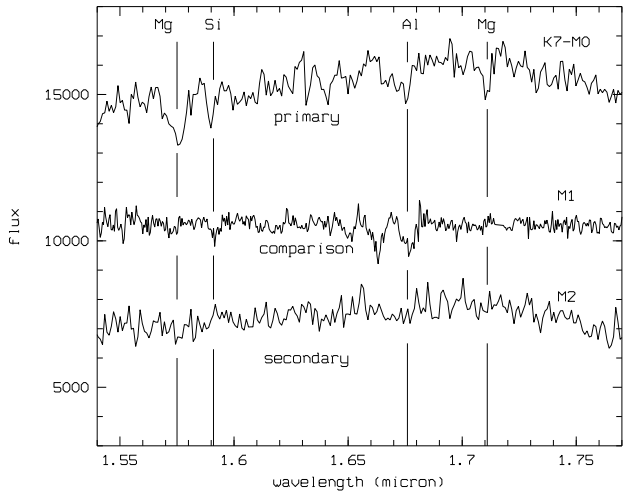,width=10.5cm,height=7.5cm,angle=270}}
\caption{H-band spectra of HD 358623 A (top) 
and B (bottom) taken with NTT/SofI
as well as and GSC~7210~1352 (in the middle) taken with VLT/ISAAC for comparison.
We spectra are not flux-calibrated; we plot a somewhat shifted flux,
so that all three objects can be compared to each other.
Star A is known to be K7-M0 with strong Mg, Si, and Al lines, 
GSC~7210~1352 is a known M1-dwarf with weaker lines.
The lines in Star B is slightly weaker than in GSC~7210~1352,
so that we classify it as M$2 \pm 1$.}
\end{figure}

As seen in Fig. 4, 
HD 358623 A shows the typical features of late-K to early-M type stars, 
namely Mg, Si, Al, and Na absorption (see e.g. Green \& Lada 1996,
Meyer et al. 1998).
Those lines are clearly weaker in the comparison star GSC~7210~1352,
a known M1 dwarf.
It is very similar for the secondary, where these lines are even a bit weaker.
Hence, regarding the presence of lines, the relative strength of them, 
and the general shape of the continuum, we can classify the secondary as 
M2 dwarf ($\pm 1$ sub-class).
This is consistent with its JHK colors (JB01 and our Table 2).
It cannot be later than M3, because a stronger than observed
absorption line appears at $1.76~\mu$m in dwarfs later than M3
(Green \& Lada 1996, Meyer et al. 1998).
This classification is also consistent with
the primary being K7-M0 and the magnitude difference between 
primary and secondary being 1.7 mag in J and H (and 1.6 mag in K),
assuming that they are at the same age and distance.

\section{Discussion and conclusion}

We have shown that the HD 358623 primary A and its companion candidate B
indeed show the same proper motion and that the spectral type of the companion 
(M2) is consistent with the observed colors and magnitude differences,
so that it is a truely bound companion.

We would like to point out again the high precision achieved in the
relative astrometry: After just one year, we could measure the proper 
motion ($\sim 100$ mas) of both HD 358623 A and B with sufficient precision
to show that they form a common proper motion pair, using the 150 mas pixel
scale of the SofI small field and several non-moving background stars.

\begin{figure}
\vspace{-3cm}
\vbox{\psfig{figure=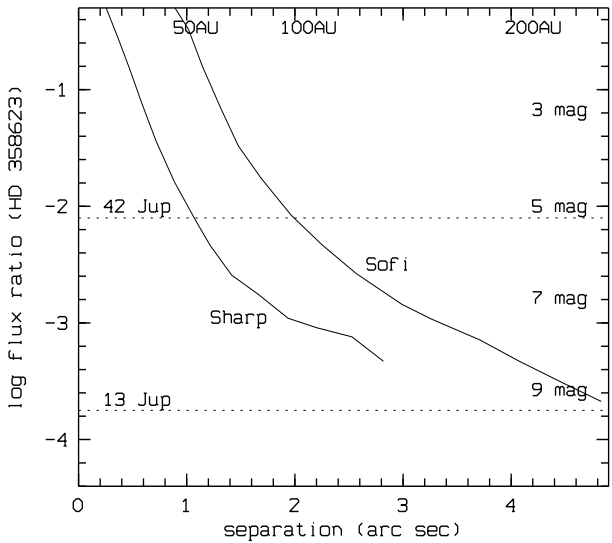,width=15cm,height=11cm,angle=270}}
\caption{Dynamic ranges achieved. We plot the log of the flux ratio between
the $3 \sigma$ background noise level and the peak intensity of HD 358623
(and, on the right hand side, the magnitude difference)
versus the separation to the primary's photocenter
(and, on the top axis, the projected physical separation at 48 pc),
for the two detectors used:
Sofi at the La Silla NTT (H-band) and SHARP-I at the NTT (K-band).
The dotted lines show the expected flux ratios for 42 and
13~M$_{\rm Jup}$ masses (according to Burrows et al. 1997) 
for 48 pc and 20 Myrs, next to HD 358623.
Every stellar companion above the H burning mass limit (75~M$_{\rm Jup}$)
would have been detected outside of $0.5^{\prime \prime}$ (24 AU at 48 pc).
The dynamic range curve for SHARP-I is limited by the 
$6^{\prime \prime} \times 6^{\prime \prime}$ size of the quadrants 
(we placed the primaries only onto the two best (lower, western) quadrants).}
\end{figure}

The mean apparent angular separation between HD 358623 A and B 
($2.205 \pm 0.028^{\prime \prime}$) corresponds to a projected
physical separation of $105.2 \pm 6.6$~AU (at the Hipparcos distance 
of HD 199143, which is presumably the same as for HD 358623, see vdA00).
The projected physical separation between HD 199143 A and B is
$48.8 \pm 3.9$~AU, namely $1.023 \pm 0.031^{\prime \prime}$ (our 
separation measured with SHARP-I) at $47.7 \pm 2.4$~pc.

Let us investigate 
the sensitivity limits determined for
the dynamic range achieved in the images:
The flux ratio is determined from our actual images of the two stars 
in all SofI and SHARP-I images as the $3 \sigma$ background noise level 
on $7 \times 7$ pixel boxes as approximate PSF areas
and devided by the peak intensity.
We compare the observed dynamic ranges with expected flux ratios for possible
companions of different masses (calculated following Burrows et al. 1997)
next to HD 358623 (Fig. 5).
The MPE speckle camera SHARP-I clearly gives the best dynamic range.
In the SHARP images, we should have detected all stellar companions
above $\sim 0.1$~M$_{\odot}$ outside of $\sim 0.5^{\prime \prime}$.
Brown dwarf companions with $\sim 25$~M$_{\rm Jup}$ would have been
detectable at $\sim 3^{\prime \prime}$ separations, more massive ones
at smaller separations (between $\sim 0.5$ and $3^{\prime \prime}$).

\begin{table}
\begin{tabular}{lccccc}
\multicolumn{6}{c}{\bf Table 4. Physical properties.} \\ \hline
Object      & T$_{\rm eff}$ & B.C. & L$_{\rm bol}$ & mass & age \\
            & [K]  & [mag] & [L$_{\odot}$]     & [M$_{\odot}$] & [Myr] \\ \hline
HD 199143 A & 6200 & 0.16  & $2.40 \pm 0.25$   & $\sim 1.25$ & $\sim 20$ \\
HD 199143 B & 3720 & 1.43  & $\sim 0.1$        & $\sim 0.60$ & $\sim 20$ \\ \hline
HD 358623 A & 3955 & 1.00  & $0.24 \pm 0.5$    & $\sim 0.90$ & $\sim 20$ \\
HD 358623 B & 3580 & 1.64  & $0.05 \pm 0.01$   & $\sim 0.55$ & $\sim 20$ \\ \hline
\end{tabular}
\end{table}

Next, we can compute the luminosities of the four objects studied.
We assume the Hipparcos distance towards HD 199143 A ($47.7 \pm 2.4$ pc) 
for all four objects.
From the known $V-H$ color indices and/or known spectral types,
we can estimate the effective temperatures and bolometric corrections 
B.C. (taken from Kenyon \& Hartmann 1995).
Temperatures, B.C., and luminosities are listed in Table 4.

We placed the stars into the H-R diagram and compared their locations
with theoretical tracks and isochrones by Palla \& Stahler (1999)
and Baraffe et al. (1998) to estimate masses and ages.
Rough values are given in Table 4.
All four stars appear to be co-eval with an age of $\sim 20$ Myrs.

In conclusion, we find that all our data are consistent with HD 199143 
and HD 358623 each having an early-M type stellar companion.
In the case of HD 358623 B, the spectral type is confirmed by 
JHK colors and a spectrum and companionship is also confirmed
by common proper motion. For HD 199143 B, spectrum as well as proper motion 
and, hence, companionship, still have to be confirmed.

\acknowledgements{
We would like to thank the NTT team with O. Hainaut, L. Vanci, and M. Billeres
for support during the SofI observations. We are gratefull to Klaus Bickert and
Rainer Sch\"odel for their help with the Sharp run.
Also, we would like to thank Wolfgang Brandner, Jo\~ao Alves, Fernando Comer\'on,
and Ray Jayawardhana for usefull discussion throughout the project.
RN wishes to acknowledge financial support from the
Bundesministerium f\"ur Bildung und Forschung through the
Deutsche Zentrum f\"ur Luft- und Raumfahrt e.V. (DLR)
under grant number 50 OR 0003. We have made use of the Simbad database 
operated at the Observatoire Strassburg.}

{}

\end{document}